\documentclass[preprint]{revtex4-1}
\usepackage{bm}
\usepackage{braket}
\usepackage{amsmath}
\usepackage{amsthm}
\usepackage{amssymb}
\usepackage{tikz}
\usetikzlibrary{positioning}
\begin{document}

\title{Advances in Synthetic Gauge Fields for Light Through Dynamic Modulation} 
\date{\today}



\author{Daniel Hey and Enbang Li}
\email[drh981@uowmail.edu.au]{}
\affiliation{School of Physics, Faculty of Engineering and Information Sciences, University of Wollongong, Wollongong, NSW 2522, Australia}

\date{\today}

\begin{abstract}
Photons are weak particles that do not directly couple to magnetic fields. However, it is possible to generate a photonic gauge field by breaking reciprocity such that the phase of light depends on its direction of propagation. This non-reciprocal phase indicates the presence of an effective magnetic field for the light itself. By suitable tailoring of this phase it is possible to demonstrate quantum effects typically associated with electrons, and as has been recently shown, non-trivial topological properties of light. This paper reviews dynamic modulation as a process for breaking the time-reversal symmetry of light and generating a synthetic gauge field, and discusses its role in topological photonics, as well as recent developments in exploring topological photonics in higher dimensions. 
\end{abstract}
	
\maketitle

\section{Introduction}

Magnetic fields are fundamental to the control of charged particles. Photons, however, are uncharged spin-1 Bosons and as a result there exists no naturally occurring gauge potential through which to control light. Manipulating the propagation of light has been a long-standing hurdle in the development of integrated photonics \cite{Huang2016, Bi2013, Doerr2014, Doerr2013, El-Ganainy2013, Bi2011}. With the development of artificial photonic structures such as photonic crystals and metamaterials \cite{Joannopoulos1997, Joannopoulos2008}, on-chip integrated photonics has been realised as not just a possibility, but increasingly more simple to achieve.

To achieve fine-control over light, one may use the physics of synthetic gauge fields. A synthetic gauge field is the tailoring of specific conditions such that some quantity of neutral particles emulate the dynamics of charged particles in a magnetic field.

\maketitle

The rotation of a trapping of neutral atoms  constitutes a synthetic gauge field: the coriolis force $\bm{F}_C = - 2 m \bm{\Omega} \times \bm{v}$ couples to the atoms in a manner analogous to the classical Lorentz force, $\bm{F}_L = q(\bm{v} \times \bm{B})$ \cite{Dalibard2015b}. Conversely, the mechanism for generating a synthetic gauge field for cold atoms relies on changing some internal degree of freedom to impart a geometric phase to the wavefunction over time \cite{Dalibard2011c}. Similar synthetic gauge fields have also been explored in optomechanics \cite{Walter2016, Ruesink2016, Yang2017} and acoustics \cite{Miri2017a, Sohn2017, Yang2016b}. This naturally lends itself to the question of whether it is possible to achieve similar conditions for light, and has spurred a great deal of research into the field of synthetic gauge fields in photonics \cite{Fang2012a, Fang2013b, Fan2015}

Symmetry is a fundamental principle in all branches of physics. The symmetry across the spatial (parity) and temporal (time) domains form the basis of what's known as parity time-symmetry (PT-symmetry), which governs how matter and energy behave under a reversal of their direction in space or time. In photonics, PT-symmetry is closely tied to the notion of reciprocity, a principle which ensures that the transfer of some quantity of light is identical between any two points regardless of geometric asymmetries in the intervening space \cite{Coulais2017}. The origin of a synthetic gauge potential for photons lies in the breaking of T-symmetry. Since reciprocity encompasses $T-$symmetry naturally, any process that breaks reciprocity will thus also break T-symmetry. The most commonly employed method of breaking T-symmetry for light is through the well-known magnetic biasing of a system, using the Faraday effect, where the magnetic field $\boldsymbol{B}$ is a pseudovector, making it odd under time-reversal \cite{Wang2006a}. However, recent techniques that bypass gyromagnetic coupling have been devised, and have opened a pathway to exploring topology in integrated photonics. 

This transfer of edifice from topological physics into photonics \cite{Haldane2008b} has created a wealth of research ideas, ranging from observations of protected edge states of light \cite{Hafezi2013, Wang2009, Poshakinskiy2014, Rechtsman2013c, Raghu2008, Barik2016} to Floquet topological insulators \cite{Chen2014, Lumer2013, Zhang2015, Leykam2016, Maczewsky2016, Khanikaev2013}. In topological photonics, spatial periodicity of a lattice is combined with a synthetic gauge field, leading to $2D$ energy bands that are characterised by the topological invariant known as the first Chern number. However, engineering of synthetic gauge fields has the potential to explore higher dimensional topologies, and the first such experimental work on $3D$ lattices has unveiled particularly intriguing topological features, including the elusive Weyl points. Going further, recent work has begun on examining $4D$ topologically non-trivial topologies of light \cite{Price2017}.

In Section \ref{t-symmetry}, we discuss the theory behind dynamic modulation as a tool for breaking T-symmetry and thus creating a synthetic gauge potential for light. We then discuss implementations of dynamic modulation and recent results in demonstrating magnetic effects for light. In Section \ref{topology}, we consider the implications of broken time-reversal symmetries for topological photonics, and examine synthetic dimensions for achieving higher dimensional topological effects.

\section{Inducing a synthetic gauge potential for light \label{t-symmetry}}



Originally, the concept of introducing a synthetic gauge field for photons was limited to a static ring resonator lattice that is carefully engineered to impart direction-dependent phases to photons with opposite spins \cite{Umucallar2011a}. On a single ring waveguide, light will resonate when it constructively interferes with itself after making a full round-trip \cite{Bogaerts2012}, in what are known as `whispering gallery' modes (so named for their acoustic origin of St Paul's cathedral, where whispered sounds propagate along the circumference of the circular interior \cite{Rayleigh1910}). For light, such modes $\omega$ correspond to the orbital angular momentum (OAM) of the mode around the ring. In a periodic array of such rings, it is possible for photons to `hop' between different rings, analogous to electrons tunnelling between atoms in a crystal. As a result the edge modes of opposite spins propagate in opposite directions, realising a photonic analogue of the quantum spin Hall effect. 

In 2009, Yu and Fan showed that interband photonic transitions induced by dynamic refractive index modulation can give rise to a gauge transformation of the photon wavefunction \cite{Yu2009a}, the precursor to what is known as dynamic modulation.

\subsection{Dynamic modulation}

\begin{figure*}
	\centering
		

	\begin{tikzpicture}[scale=1,every node/.style={minimum size=1cm},on grid]
	\large
	\draw[step=2cm,color=black] (-3,-3) grid (3,3);
	\draw[fill=red] (-2,-2) circle (0.2cm);
	\draw[fill=blue] (0,-2) circle (0.2cm);
	\draw[fill=red] (2,-2) circle (0.2cm);
	
	\draw[fill=blue] (-2,0) circle (0.2cm);
	\draw[fill=red] (0,0) circle (0.2cm);
	\draw[fill=blue] (2,0) circle (0.2cm);
	
	\draw[fill=red] (-2,2) circle (0.2cm);
	\draw[fill=blue] (0,2) circle (0.2cm);
	\draw[fill=red] (2,2) circle (0.2cm);
	
	\draw[thick,->] (-2.7,1.2) -- (-2.7,2);
	\draw[thick,->] (-2.7,0.8) -- (-2.7,0);
	
	\node at (-2.7,1) {\large $a$};
	\node at (-3.5,2) {\large $\omega_A$};
	\node at (-3.5,0) {\large $\omega_B$};
	
	\node at (-1.6,1) {$\phi_{ij}$};
	\end{tikzpicture}
	
	\caption{\textbf{a)} A lattice of photonic resonators, with two square sub-lattices of resonators with frequency $\omega_A$ (red) and $\omega_B$ (blue), respectively. There is only nearest-neighbour dynamic coupling \cite{Fang2012a}.}
	\label{}
\end{figure*}
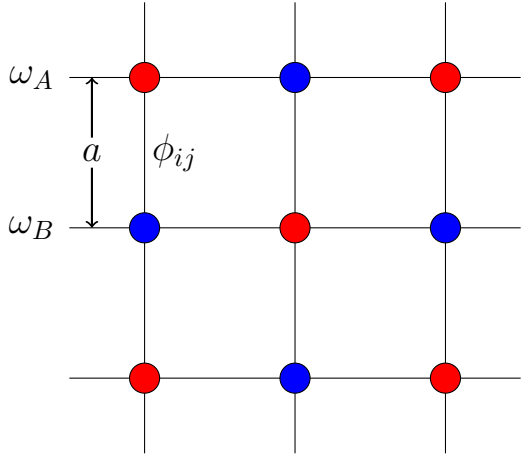

Generating a synthetic gauge field for light relies on dynamic modulation of inter-penetrating photonic lattices. Here, we shall briefly discuss the general theory behind such modulation following the proposal of Fang et al. \cite{Fang2012a}. Their model consists of a lattice of two inter-penetrating resonators of frequencies $\omega_A$, $\omega_B$ and nearest neighbour separation (lattice constant) $a$. We shall assume nearest neighbour coupling is achieved by a time-dependent harmonic index modulation of frequency $\Omega$ and coupling strength $V$. In the tight-binding model, the Hamiltonian of such a system is given by

\begin{equation}
	H = \omega_A \sum_{i} a^{\dagger}_i a_i + \omega_B \sum_{j} b^{\dagger}_j b_j + \sum_{<ij>} V \cos{(\Omega t + \phi_{ij})}(a^{\dagger}_i b_j + b^{\dagger}_j a_i)
\end{equation}

for creation operators on the lattice $a\dagger,b\dagger$ \cite{Fang2012} respectively. Now, it is assumed that the modulation is on resonance such that the modulation frequency $\Omega = \omega_B - \omega_A$. For a coupling strength $V \leq \Omega$, the rotating-wave approximation can be applied to transform the Hamiltonian of the system into 

\begin{equation}
H = \sum_{<ij>} \dfrac{V}{2} (e^{i \phi_{ij}} c^{\dagger}_i c_j+e^{i\phi_{ij}} c^{\dagger}_j c_i),
\end{equation}

where $c_{i(j)} = e^{i \omega_{A(B)} t} a_i (b_j)$. 
This is identical to the Hamiltonian of a charged particle in a periodic lattice subject to a magnetic field \cite{Luttinger1951}, so long as we identify

\begin{equation}
\phi_{ij} = \int_{i}^{j} \bm{A}_{eff} d\bm{l}. 
\end{equation}

This modulation phase, $\phi_{i,j}$, gives rise to a direction-dependent hopping term between nearest neighbours, thus generating an \textit{effective} gauge potential for photons on the lattice, as in Figure 1. This forms the origin of the synthetic gauge field for photons on a lattice undergoing dynamic modulation. This phase is strictly non-reciprocal, and consequently, possesses a broken T-symmetry; reverting the direction of propagation will reverse the sign of phase.

The phase acquired by the photon as it undergoes dynamic modulation imposes a gauge potential for light. As far as the photon is concerned, the phase of some wavefunction cannot be directly measured, only differences in phase, and thus represents a gauge degree of freedom \cite{Li2014a}. This phase possesses the same gauge ambiguity as that of the phase acquired in an electronic gauge potential along an open path. Likewise, reverting the direction of propagation results in a change of sign. Consequently, this phase has the same properties as the electronic Aharonov-Bohm phase, so that demonstrating the non-reciprocity is equivalent to demonstrating the existence of a gauge potential for photons. As an aside, we note that several experiments have successfully observed Aharonov-Bohm phases for light \cite{Li2014a, Tzuang2014, Fang2013c}.

Dynamically modulated resonator lattices also support one-way edge modes. Since the Hamiltonian describing the system is temporally periodic, the concepts of Floquet's theorem can be used to illustrate the band structure. The resulting Floquet band structure has the same gapless edge states as that of a static quantum Hall phase.

However, time-harmonic modulations of more than a few resonators simultaneously is a challenging feat at optical wavelengths \cite{Tzuang2014}. It has been shown however by Rechtsman et al. that it is possible to shift the modulation from the frequency to the spatial domain  to observe photonic analogues of the quantum Hall effect at optical frequencies \cite{Rechtsman2013b}, corresponding to the first experimentally viable Floquet topological phases \cite{Lindner2011}. To achieve this, Rafezi et al. extended the periodic lattice along $z$, such that $z$ plays the role of time. Breaking the $z$ symmetry of the system is then analogous to breaking $T$ symmetry of a regular system, which opens up band degeneracies in the Floquet bandstructure - culminating in protected edge modes.

Finally, we note that most previous works have focused solely on the regime where the modulation strength is far less than the modulation frequency \cite{Lin2014, Lee2017a, Tzuang2014, Fang2012}, allowing for the application of the \textit{rotating-wave approximation}. On the other hand, there has been recent interest in light-matter interactions in the \textit{ultra-strong coupling regime} \cite{Fedortchenko2016, Hirokawa2017, Sanchez-Burillo2015, Irish2005, Rabl2011, Schiro2012, Sanchez-Burillo2014}, where the RWA is no longer valid. In such ultra-strong coupling systems, it has been shown that topologically protected one-way edge states in dynamic modulation are less susceptible to intrinsic losses \cite{Fan2015}. In addition to this, a unique topological phase transition was found to be associated with variations in the modulation strength \cite{Yuan2015a}. Such phase transitions are not found in the rotating-wave counterparts. 

\definecolor{red1}{RGB}{226,56,56}
\definecolor{orange1}{RGB}{247,130,0}
\definecolor{yellow1}{RGB}{255,185,0}
\definecolor{green1}{RGB}{94,189,62}
\definecolor{blue1}{RGB}{0,156,223}
\definecolor{grey1}{RGB}{195,195,195}

\section{Topological photonics \label{topology}}

When a two-dimensional electron gas at low temperature is placed in an out-of-plane magnetic field, the transverse conductance of the gas becomes quantized, a phenomenon known as the 2D quantum Hall effect \cite{Tong2016}. \textit{Topology} arises as a macroscopic quantity of a system. Suppose the external field is applied along such that $\bm{E} = E_y e^y$, then the conductance

\begin{equation}
j^x = -eE_y \nu_1
\end{equation}

where $e$ is the electron charge, and $\nu_1$ is a topological invariant known as the first Chern number. 

\begin{equation}
\nu_1 = \dfrac{1}{2\pi} \int_{BZ} \mathcal{F}^{xy} d^2 k 
\end{equation}

Where the integral occurs over the Brillouin zone (BZ) and $\mathcal{F}$ is the Berry curvature,

\begin{equation}
\mathcal{F}^{xy} = i\Big[\braket{\dfrac{\partial u}{\partial k_x}|\dfrac{\partial u}{\partial k_y}} - \braket{\dfrac{\partial u}{\partial k_y}|\dfrac{\partial u}{\partial k_x}} \Big]
\end{equation}

and $\ket{u(\bm{k})}$ are the Bloch states of the energy band \cite{Price2016a}.
A topological invariant is a quantity that remains constant under any kind of deformation of a system. Consequently, the first Chern number cannot be changed by small perturbations or impurities in the system. For charged particles, the most practical way of engineering a non-zero Chern number is through application of a magnetic flux. 

On the other hand, Haldane's model in the late 1980's showed that it is possible to observe a quantum Hall effect without external magnetic fields, but rather through broken T-symmetry \cite{Haldane1988}. Such topological devices with broken T-symmetry are known as Floquet insulators. One of the most noticeable hallmarks of Floquet insulators is the presence of robust gapless edge modes \cite{Hasan2010}.

\subsection{Synthetic dimensions}

\begin{figure*}
	\centering
	\begin{tikzpicture}[scale=1.5,every node/.style={minimum size=1cm},on grid]
	\large
	\draw[fill=grey1] (2.5,2.5) circle (2.3cm);
	\draw[fill=white] (2.5,2.5) circle (2.1cm);
	\draw[thick, ->] (0.5,2.5) arc (180:240:2cm);
	\draw[thick, ->] (3.5,0.8) arc (300:360:2cm);
	\draw[black,very thick,fill=white] (3,0) rectangle (2,0.5);
	
	\draw[thick,<->] (6,4.2) -- (6,3.8);
	\draw[thick,<->] (6,3.2) -- (6,2.8);
	\draw[thick,<->] (6,2.2) -- (6,1.8);
	\draw[thick,<->] (6,1.2) -- (6,0.8);

	\draw[fill=blue1] (6,4.5)  circle (0.3cm);
	\draw[fill=green1] (6,3.5)  circle (0.3cm);
	\draw[fill=yellow1] (6,2.5)  circle (0.3cm);
	\draw[fill=orange1] (6,1.5)  circle (0.3cm);
	\draw[fill=red1] (6,0.5)  circle (0.3cm);
	
	\node at (7,4.5) {$\omega_0+2\Omega$};
	\node at (7,3.5) {$\omega_0+\Omega$};
	\node at (7,2.5) {$\omega_0$};
	\node at (7,1.5) {$\omega_0-\Omega$};
	\node at (7,0.5) {$\omega_0-2 \Omega$};
	\node at (2.5,0.25) {$\Omega_M$};
	
	\foreach \x in {4.5,...,0.5}
	\draw[fill=white] (6,\x) circle (0.25cm);
	\end{tikzpicture}
	\caption{A ring resonator, which supports resonant modes, undergoing dynamic modulation by an electro-optic modulator. The resonator modes form a synthetic dimension along the frequency axis $\omega$.}
	\label{}
\end{figure*}
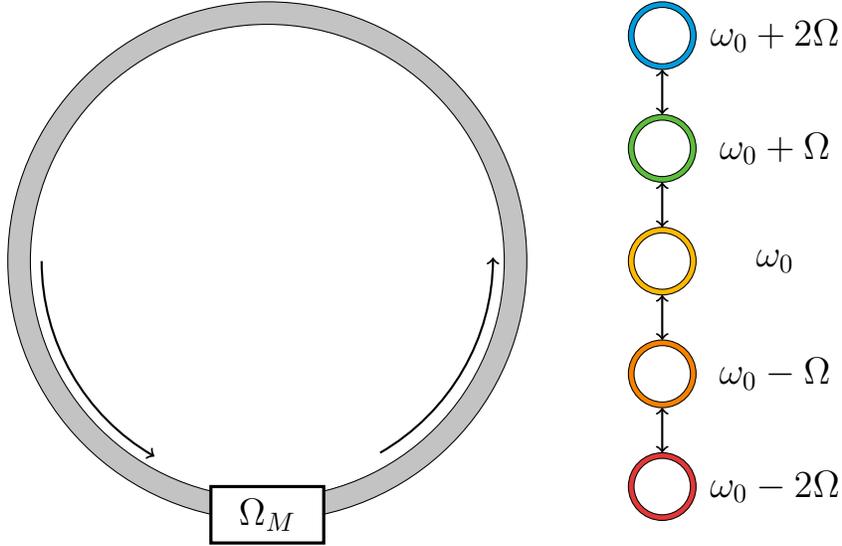
Studies in topology have long hinted at rich possibilities of physics in higher dimensions, namely analogues of the quantum Hall effect in even-dimensional spaces63. Traditional systems in condensed matter physics however, are locked out of examining such effects. Only recently have new approaches in simulating higher dimensional topological models, so-called synthetic dimensions have been proposed, that generate a gauge field for neutral particles in a synthetic dimensio \cite{Celi2014, Saito2017, Price2017, Barbarino2016, Mancini2015, Stuhl2015a, Mancini2015}. These ideas were first proposed in ultracold atomic gases, where the atoms possess an internal spin degree of freedom. Photons naturally possess many internal degrees of freedom: frequency, orbital angular momentum, spin angular momentum, polarization, and so on. These additional degrees of freedom can form synthetic lattice dimensions for light, leading to possible experimental analysis of higher dimensional topological photonics.

Here we demonstrate a basic method of achieving synthetic dimensions in dynamic modulation. Consider one of the most elementary photonic structures: the ubiquitous ring resonator. In the absence of group velocity dispersion (GVD), the single ring supports a set of  equally spaced resonant frequencies. The spacing of the modes is based on the spectral range of the resonator, and is related to the round trip time $T$ by $\Omega = {2 \pi}/{T}$.

By introducing an electro-optic phase modulator on the ring, as shown in Figure 2, it is possible to generate side-bands. If the modulation frequency $\Omega_M$ is then chosen to be equal to the mode spacing of the resonator $\Omega$,there will be a resonant coupling of the modes that are separated by the free spectral range of the ring. Consequently, this system can be described by a $1D$ tight-binding model, despite being a $0D$ structure. Of course, the additional dimension is the frequency dimension.   

This can naturally be extended to the case of an array of ring resonators, as in Figure 3, that are coupled together to form a waveguide \cite{Yuan2016c}. Each ring has a controlled phase of electro-optic modulation, which gives rise to a $2D$ physics model on a $1D$ structure. The phase of the modulation corresponds then to the hopping phase along the frequency axis.

It is possible to generate a topologically non-trivial bulk correspondence by considering boundaries in real and synthetic space. In real space, the boundary is given by the physical edges of the ring. However, in frequency space, a boundary is introduced by the group velocity dispersion instead. Around the zero-GVD point, most frequencies are equally spaced so the modulators induce an on-resonance coupling. On the other hand, away from the zero-GVD point, the frequencies are no longer equally spaced and thus can not support coupling, leading to an effective boundary in frequency space. 

Finally, we note that outside of applications in topological photonics discussed below, synthetic dimensions in dynamic modulation can also lead to new possibilities for controlling the frequencies of light \cite{Qin2017a}. 

\begin{figure*}
	\centering
	\begin{tikzpicture}[scale=1.5,every node/.style={minimum size=1cm},on grid]
	\large
	\begin{scope}[
	yshift=0]
	\draw[fill=grey1] (2.5,2.5) circle (1.2cm);
	\draw[fill=white] (2.5,2.5) circle (1.1cm);
	
	\draw[fill=blue1] (2.5,0.5)  circle (0.2cm);
	\draw[fill=green1] (2.5,-0.5)  circle (0.2cm);
	\draw[fill=yellow1] (2.5,-1.5)  circle (0.2cm);
	\draw[fill=orange1] (2.5,-2.5)  circle (0.2cm);
	\draw[fill=red1] (2.5,-3.5)  circle (0.2cm);
	
	\draw[thick,<->] (2.8,0.5) -- (5.2,0.5);
	\draw[thick,<->] (2.8,-0.5) -- (5.2,-0.5);
	\draw[thick,<->] (2.8,-1.5) -- (5.2,-1.5);
	\draw[thick,<->] (2.8,-2.5) -- (5.2,-2.5);
	\draw[thick,<->] (2.8,-3.5) -- (5.2,-3.5);
	
	\node at (4,0.8) {$V$};
	
	\foreach \x in {0.5,...,-3.5}
	\draw[fill=white] (2.5,\x) circle (0.15cm);
	\begin{scope}[yshift=-4cm]
	\node at (1.5,4.5) {$\omega_0+2\Omega$};
	\node at (1.5,3.5) {$\omega_0+\Omega$};
	\node at (1.5,2.5) {$\omega_0$};
	\node at (1.5,1.5) {$\omega_0-\Omega$};
	\node at (1.5,0.5) {$\omega_0-2 \Omega$};
	\end{scope}
	
	\begin{scope}[yshift=-4cm]
	\draw[thick,<->] (2.5,4.2) -- (2.5,3.8);
	\draw[thick,<->] (2.5,3.2) -- (2.5,2.8);
	\draw[thick,<->] (2.5,2.2) -- (2.5,1.8);
	\draw[thick,<->] (2.5,1.2) -- (2.5,0.8);
	
	\end{scope}
	\node at (2.5,1.7) {$\phi_1$};
	
	\end{scope}
	
	\begin{scope}[
	xshift=3cm]
	\draw[fill=grey1] (2.5,2.5) circle (1.2cm);
	\draw[fill=white] (2.5,2.5) circle (1.1cm);
	
	\draw[fill=blue1] (2.5,0.5)  circle (0.2cm);
	\draw[fill=green1] (2.5,-0.5)  circle (0.2cm);
	\draw[fill=yellow1] (2.5,-1.5)  circle (0.2cm);
	\draw[fill=orange1] (2.5,-2.5)  circle (0.2cm);
	\draw[fill=red1] (2.5,-3.5)  circle (0.2cm);
	
	\foreach \x in {0.5,...,-3.5}
	\draw[fill=white] (2.5,\x) circle (0.15cm);
	\node at (2.5,1.7) {$\phi_2$};
	
	\draw[thick,<->] (2.8,0.5) -- (5.2,0.5);
	\draw[thick,<->] (2.8,-0.5) -- (5.2,-0.5);
	\draw[thick,<->] (2.8,-1.5) -- (5.2,-1.5);
	\draw[thick,<->] (2.8,-2.5) -- (5.2,-2.5);
	\draw[thick,<->] (2.8,-3.5) -- (5.2,-3.5);
	
	\begin{scope}[yshift=-4cm]
	\draw[thick,<->] (2.5,4.2) -- (2.5,3.8);
	\draw[thick,<->] (2.5,3.2) -- (2.5,2.8);
	\draw[thick,<->] (2.5,2.2) -- (2.5,1.8);
	\draw[thick,<->] (2.5,1.2) -- (2.5,0.8);
	\end{scope}
	\end{scope}
	
	\begin{scope}[
	xshift=6cm]
	\draw[fill=grey1] (2.5,2.5) circle (1.2cm);
	\draw[fill=white] (2.5,2.5) circle (1.1cm);
	
	\draw[fill=blue1] (2.5,0.5)  circle (0.2cm);
	\draw[fill=green1] (2.5,-0.5)  circle (0.2cm);
	\draw[fill=yellow1] (2.5,-1.5)  circle (0.2cm);
	\draw[fill=orange1] (2.5,-2.5)  circle (0.2cm);
	\draw[fill=red1] (2.5,-3.5)  circle (0.2cm);
	
	\foreach \x in {0.5,...,-3.5}
	\draw[fill=white] (2.5,\x) circle (0.15cm);
	\node at (2.5,1.7) {$\phi_3$};
	
	\begin{scope}[yshift=-4cm]
	\draw[thick,<->] (2.5,4.2) -- (2.5,3.8);
	\draw[thick,<->] (2.5,3.2) -- (2.5,2.8);
	\draw[thick,<->] (2.5,2.2) -- (2.5,1.8);
	\draw[thick,<->] (2.5,1.2) -- (2.5,0.8);
	\end{scope}
	\end{scope}
	
	\end{tikzpicture}
	\caption{Extending the 0D ring-resonator of Figure 1 into an array of equally spaced resonators in 1D allows for the study of 2D physics on a 1D structure, by again exploiting the transitions between frequencies.}
	\label{3DDynamicLattice}
\end{figure*}
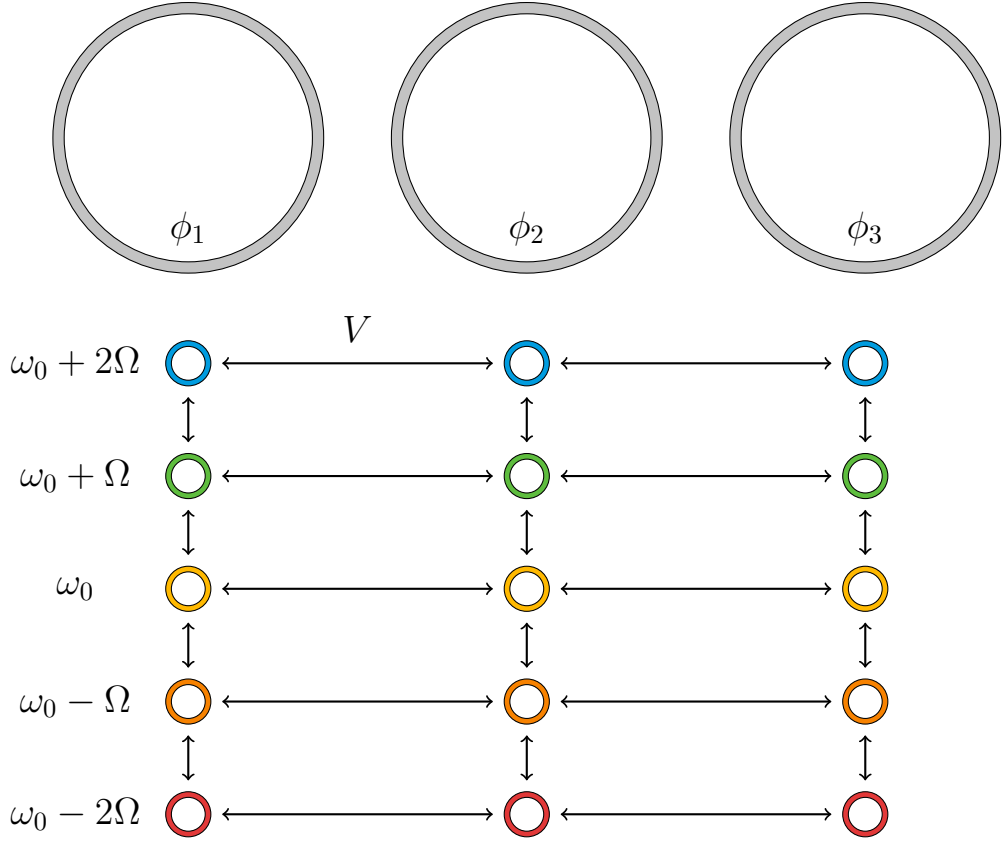

\subsection{Weyl points}

In condensed matter systems, the Weyl point describes a magnetic monopole in momentum space, that is, a `source' or `sink' of Berry curvature. As such, the Weyl point is perceived as topological nodal points in $3D$ momentum space. Weyl points are topologically robust, in the sense that they can not be destroyed by any perturbation that preserves translational symmetry. The Hamiltonian of a Weyl system is
\begin{equation}
H = k_x \sigma_x + k_y \sigma_y + k_z \sigma_z + E_0 I,
\end{equation}
where $\sigma_{x,y,z}$ are the Pauli spin matrices, and $I$ is the $2x2$ identity matrix, which together form a complete basis for $2x2$ Hermitian matrices. When only P-symmetry is broken, the minimum number of pairs of Weyl points is two because T maps a Weyl point at $k$ to $-k$ without changing its Chern number. When only T-symmetry is broken, the minimum number of pairs of Weyl points is one. There has been a massive effort devoted to investigating Weyl points and their associated phenomena in electronic systems, including Fermi arc surface states \cite{Wu2016, Huang2015, Xu2015, Gupta2017} and chiral anomalies \cite{Lv2015, Zhou2015, Burkov2014, Zyuzin2012, Spivak2016, Gorbar2014}. Weyl points have also been found in photonic crystals \cite{Chang2017, Noh2017, Bravo-Abad2015a, Xiao2016, Lu2013, Wang2016b, Chen2016a}, as well as acoustic \cite{Yang2016a, Xiao2015} and plasmonic structures \cite{Gao2016a}.

To explore a Weyl point in a planar 2D geometry, one may use a synthetic dimension \cite{Celi2014} to simulate the third spatial dimension. The size of the synthetic dimension, which corresponds to the number of modes in each individual ring can be chosen to be almost arbitrarily large without increasing the system complexity. In a recent paper, Lin et al suggested creating a synthetic $3D$ space by dynamically modulating a $2D$ array of on-chip ring resonators \cite{Lin2017}. Each resonator supports a set of discrete modes equally spaced in resonant frequency space. These discrete modes thus form a periodic lattice in the third, synthetic frequency dimension. The two spatial dimensions and one synthetic frequency dimension together form a $3D$ space. Their proposed approach is specifically designed for implementation using an existing on-chip integrated photonic platform. Compared with previous complex electromagnetic or acoustic structures for realizing Weyl points \cite{Bravo-Abad2015, Lu2015, Rechtsman2013, Gao2016, Chen2016}, a planar 2D approach is simpler and more flexible, and can be achieved with the previously discussed extension of dynamic modulation in synthetic dimensions, allowing for on-chip investigation of Weyl points. By changing the phases of dynamic modulation, the device can be tuned to exhibit Weyl points under a breaking or P or T symmetry. 

\subsection{Towards 4D photonics}

Naturally, the quantum Hall effect was first generalized to higher dimensions for electronic systems \cite{Zhang2001, Kraus2013, Hasebe2014, Mathai2015, Edge2012}. 
It is a standard extension of our discussion on $2+1D$ photonic Weyl points then to consider a \textit{3 dimensional} array of resonators in the three spatial dimensions $x, y, z$, and then extending it into the synthetic frequency dimensions $\omega$ to probe elusive 4D quantum Hall effects.

A 4-dimensional model making use of 3D resonating lattices with plaquettes has only recently been proposed by Price et al. \cite{Price2016a}, based on the authors previous works in 4D quantum Hall effects for ultra-cold atoms \cite{Price2015}. This model differs to a potential synthetic dimensions with dynamic modulation, as the lattice their model does not break $T-$symmetry. Because of this, photons with positive angular momentum can backscatter into states with a negative angular momentum. However, it has been noted that such backscattering can be minimised during fabrication \cite{Hafezi2013}. Finally, we note that achieving a $4D$ topological effect for light would not only be possible under a dynamic modulation regime, but also would involve a broken T-symmetry, leading to one-way edge modes propagating along $3D$ frequency space.

\section{Conclusion and Outlook}
Photonic circuits are currently hampered by imperfections in the manufacturing process, so a method to bypass flaws via synthetic gauge fields would be hugely significant. Exploitation of topological effects dramatically improves the robustness of current photonic devices, including optical circulators and isolators, which currently require magneto-optic coupling effects, and are largely incompatible with integrated photonics. Likewise, the topologically non-trivial properties of light could solve key limitations involving the disorder of light in waveguides, and potentially improve coherence in quantum computing.  In-fact, protected edge states arising from T-invariant topological phases would be one-way, even in reciprocal systems, making current optical isolators unnecessary. The main challenge in dynamic modulation schemes mostly lies within engineering difficulties associated with modulating coupled resonators on the micro-scale. 

Going further, the recent introduction of synthetic dimensions is potentially an important step towards realising higher-dimensional topology in photonics. The investigation of Weyl points on a 2D planar structure makes it far more experimentally viable than current architectures, and is amenable with integrated photonics in both the infrared and visible wavelengths. The idea of Weyl points in 2D systems with synthetic dimensions can be implemented in more than just honeycomb lattice systems. Additionally, a 4D quantum hall effect proposed, based on a $3D$ coupled resonator system with an additional frequency dimension \cite{Price2016a}, allows for unique photon-photon interactions in $3D$ systems, and could potentially realise fractional quantum hall physics for light. Exciting applications of these approaches would obviously be the observation of a $4D$ quantum Hall effect experimentally. Further, the edge physics associated with a 4D Hall effect would expect to propagate along the $3D$ surface of a $4D$ boundary. Such modes would have useful applications in future optical components.

The mutually beneficial relationship between fields of research in ultra-cold atoms and optical synthetic gauge fields is important, owing to the fact that developments in topological photonics can be easily translated to other bosonic systems. Just as topological insulators in electronic systems have pushed forward the limits of fundamental physics, topological photonics promises to catalyse a new era of optical devices.

\vskip1pc

\pagebreak

	\bibliographystyle{RS}
	\bibliography{library}

\end{document}